\documentstyle[amsmath,amssymb,epsf,mncite]{mn}\voffset=-.5in
\def\bildchenwidth{.707\linewidth}

\def\bildchen#1#2{\begin{minipage}{#1}\epsfxsize=#1\epsfbox{#2}\end{minipage}}

\def\<{\ifmmode\langle \else$\langle$\fi}
\def\>{\ifmmode\rangle \else$\rangle$\fi}

\def\deg{\ifmmode{^\circ}\else{$^\circ$}\fi}
\def\hGpc{\ifmmode{h^{-1}{\rm Gpc}}\else{$h^{-1}{\rm Gpc}$}\fi}
\def\hMpc{\ifmmode{h^{-1}{\rm Mpc}}\else{$h^{-1}{\rm Mpc}$}\fi}

\def\muK{\ifmmode{\mu{\rm{K}}}\else{$\mu$K}\fi}
\def\mum{\ifmmode{\mu{\rm{m}}}\else{$\mu$m}\fi}

\def\Eins{{\bf 1}}

\def\bigdcup{\mathop{\bigcup\mkern-16mu\cdot\kern-2.77779pt\mkern16mu}}
\def\binko#1#2{\left(\!\begin{array}{c}{#1}\\{#2}\end{array}\!\right)}
\newlength{\epsfrest}

\def\grad{\mathop{\bf grad}}

\def\tfrac#1#2{{\textstyle\frac{#1}{#2}}}

\def\da{{{\rm d}a}}

\def\dl{{{\rm d}\ell}}

\def\ds{{{\rm d}s}}
\def\dt{{{\rm d}t}}

\def\dv{{{\rm d}v}}

\def\bu{{\bf u}}

\def\bx{{\bf x}}

\def\bsigma{{\boldsymbol{\sigma}}}

\def\cI{{\mathcal I}}

\def\gM{{\mathfrak M}}
\def\gS{{\mathfrak S}}

\def\rd{{\rm d}}

\def\EE{{\mathbb{E}}}
\def\HH{{\mathbb{H}}}

\def\Ss{{\mathbb{S}}} 
\def\ups{\upsilon}

\pubyear{1997}
\volume{000}
\pagerange{\pageref{firstpage}--\pageref{lastpage}}
\begin{document}

\label{firstpage}

\title[Minkowski Functionals of CMB Maps]{ 
Minkowski Functionals used in the Morphological Analysis
of Cosmic Microwave Background Anisotropy Maps }

\author[Jens Schmalzing and Krzysztof M.\ G{\'o}rski]{
Jens Schmalzing$^{1,2,\star}$ and Krzysztof M.\ G{\'o}rski$^{3,4,\dag}$
\vspace*{1ex}\\
$^1$Max--Planck--Institut f\"ur Astrophysik,
Karl--Schwarzschild--Stra{\ss}e 1,
85740 Garching, Germany
\\
$^2$Ludwig--Maximilians--Universit\"at,
Theresienstra{\ss}e 37,
80333 M\"unchen, Germany
\\
$^3$Theoretical Astrophysics Center,
Juliane Maries Vej 30,
2100 K{\o}benhavn \O, Denmark
\\
$^4$Warsaw University Observatory,
Aleje Ujazdowskie 4,
00--478 Warszawa, Poland
\\
$^\star$email jensen@mpa-garching.mpg.de
\\
$^\dag $email gorski@tac.dk
}

\date{Version of 20 December 1997.  Accepted for publication in Monthly Notices.}

\maketitle

\begin{abstract}
We  present a novel approach  to  quantifying the morphology of Cosmic
Microwave  Background (CMB)     anisotropy  maps.   As   morphological
descriptors,  we use shape parameters  known as Minkowski functionals.
Using the mathematical  framework provided by  the  theory of integral
geometry on arbitrary curved supports, we point out the differences to
their characterization  and interpretation in the  case of flat space.
With restrictions of real data  -- such as pixelization and incomplete
sky coverage, to  mention just a  few -- in  mind, we derive and  test
unbiased estimators for  all Minkowski functionals.  Various examples,
among  them the analysis of  the  four--year COBE DMR data, illustrate
the application of our method.
\end{abstract}

\begin{keywords}
{methods: numerical; methods: statistical; cosmic microwave background}
\end{keywords}

\section{Introduction}

%
The oldest signal accessible   to  mankind  is the Cosmic    Microwave
Background discovered by {}\scite{penzias:measurement}.  Consisting of
photons that  have been  free--streaming since  the  Universe was only
300,000 years old,    the Cosmic Microwave Background  (CMB)  provides
valuable information on the early history  of our Universe.  Above all
its  anisotropies mirror  the   matter fluctuations  at the   epoch of
recombination (at redshift $z\approx1,100$) and with them the seeds of
the large--scale structure seen today.

Various   methods of statistical   analysis  have been used on  Cosmic
Microwave Background maps.  Among them are  the two-- and three--point
correlation           function            ({}\pcite{hinshaw:twopoint},
{}\pcite{hinshaw:threepoint}),            the      power      spectrum
{}\cite{gorski:power},  skewness   and kurtosis {}\cite{luo:kurtosis},
multifractals    {}\cite{pompilio:multifractal},   and   the   extrema
correlation  function     {}\cite{kogut:gaussian}.   Another promising
approach is the investigation of the morphology of hot and cold spots.
Being  complementary to the traditional  approach via the hierarchy of
correlation functions, it provides alternative methods for determining
cosmological    parameters    {}\cite{torres:genus}.   But  above all,
morphological  statistics   incorporate  correlation   functions    of
arbitrary   order.   Hence    they  are  sensitive  to signatures   of
non--Gaussianity in the temperature fluctuations, which would indicate
the  presence  of  topological defects such   as   strings or textures
arising   from    phase  transitions     in    the    early   Universe
{}\cite{stebbins:cosmic}.

In  order to  measure  the morphology  of Cosmic Microwave  Background
anisotropies, the Euler characteristic, or equivalently the genus, was
suggested as    long as  a  decade  ago  ({}\pcite{coles:nongaussian},
{}\pcite{coles:statistical}).   Even  to  date,  most applications are
confined  to genus statistics, although  an early theoretical study by
{}\scite{gott:topologymicrowave} also considers the boundary   length,
but failed  to  come up  with a  subsequent   analysis of data.    The
analysis of the first--year COBE  DMR data using genus statistics  was
done  by  {}\scite{smoot:topologyfirst};  their work  also  contains a
thorough discussion of the performance of the method compared to other
measures  of  non--Gaussianity.  Further   applications of topological
methods on CMB anisotropies come from {}\scite{torres:topological} and
{}\scite{torres:genus}, and genus  calculations of the four--year COBE
DMR       data   are     due      to    {}\scite{colley:topology}  and
{}\scite{kogut:tests}.

The genus can  be  placed in   the  wider framework of   the Minkowski
functionals  {}\cite{minkowski:volumen},  by   natural and  compelling
mathematical considerations.  Originally introduced to tackle problems
of  stochastic  geometry, this   family  of morphological  descriptors
subsequently set  off    the development of  integral   geometry  (see
{}\pcite{blaschke:I} or {}\pcite{hadwiger:vorlesung} for early  works,
and   {}\pcite{schneider:brunn}  for   a   comprehensive    overview).
Recently,  the   Minkowski functionals   have   been introduced   into
cosmology    as descriptors    for the    morphological  properties of
large--scale  structure  by   {}\scite{mecke:robust}.  While     their
original  approach uses a  Boolean    grain model applicable  to   the
analysis of point sets, {}\scite{schmalzing:beyond} consider excursion
sets and isodensity contours of  smoothed random fields.  Applications
have   so  far been restricted    to  the morphometry  of large--scale
structure      in     redshift      catalogues       of       galaxies
{}\cite{kerscher:fluctuations}   and   clusters     of        galaxies
{}\cite{kerscher:abell}.

The promising results  in large--scale structure analysis motivate the
application of  Minkowski  functionals to  Cosmic Microwave Background
sky maps.   Since all--sky maps live on  a curved support, some formal
obstacles will be encountered,  but the underlying concepts remain the
same,  and    the    central  formulae     are  easily     generalized
{}\cite{santalo:integralgeometry}.  In applications to data, care must
be taken  to  remove the effects  of usually  incomplete sky coverage,
while retaining as much information as possible at the same time.

Our article   is  organized    as follows.    Section~\ref{sec:theory}
summarizes the framework of integral geometry first in flat space, and
then on  spaces  of  non--zero  but constant curvature,   with special
regard to the   different  interpretations.   Also, we   express   all
Minkowski functionals of the excursion set of a smooth random field as
integrals over purely local  invariants  formed from derivatives.   In
Section~\ref{sec:minkowski} we  use   these   integration formulae  to
construct estimators  for  the numerical evaluation   of the Minkowski
functionals for   a pixelized CMB sky map.    The important problem of
incomplete sky coverage is addressed,  and  we find prescriptions  for
boundary corrected,   unbiased  estimators, even  for  smoothed  data.
Section~\ref{sec:examples} is devoted to examples,  among them a study
of  noise reduction through  Gaussian  filtering, and a  morphological
analysis of the maps constructed from  the COBE DMR four--year data by
{}\scite{bennett:fouryeardata}.  Finally,  we summarize,  draw     our
conclusions and provide an outlook  in Section~\ref{sec:outlook}.  Two
appendices further illuminate  the mathematical aspects of  this paper
by giving detailed derivations of important formulae.

\section{Theory}
\label{sec:theory}

\subsection{Integral geometry}

\begin{table}
\begin{center}\begin{tabular}{c|ccc}
$d$ & 1 & 2 & 3 \\
\hline
$V_0$ & length & area          & volume \\
$V_1$ & $\chi$ & circumference & surface area \\
$V_2$ & --     & $\chi$        & total mean curvature \\
$V_3$ & --     & --            & $\chi$ \\
\end{tabular}\end{center}
\caption{
Some of the  $d+1$ Minkowski functionals in $d$--dimensional Euclidean
space may be interpreted  as familiar geometric quantities (apart from
numerical   factors).    This   table    summarizes    the   geometric
interpretations of all Minkowski  functionals  for one, two and  three
dimensions.  The symbol $\chi$   denotes the Euler  characteristic,  a
purely topological quantity; it  measures  the connectivity of a  set,
being equal to  the number of  parts minus the number  of holes in two
dimension.
\label{tab:geometric}
}
\end{table}

Let us first introduce integral geometry in flat space, or, to be more
precise, in $d$--dimensional  Euclidean   space $\EE^d$.  We wish   to
characterize  the   morphology  of a  suitable  set $Q\subseteq\EE^d$.
Hadwiger's Theorem {}\cite{hadwiger:vorlesung} states that under a few
simple  requirements,  any    morphological descriptor   is  a  linear
combination  of only   $d+1$  functionals; these  are the   so--called
Minkowski functionals $V_j$, with  $j$ ranging from 0  to $d$.  If the
set $Q$ has a smooth boundary $\partial{Q}$, its Minkowski functionals
--  except for the  $d$--dimensional volume $V_0$,  which is of course
calculated   by volume integration  --    are given by  simple surface
integrals  {}\cite{schneider:curvature}.     So      altogether     we
have\footnote{We use  $\omega_j$ to  denote  the surface  area  of the
$j$--dimensional unit sphere.   Some special values  are $\omega_0=2$,
$\omega_1=2\pi$, $\omega_2=4\pi$, while in general
\begin{equation}
\omega_{j}=\frac{2\pi^{(j+1)/2}}{\Gamma((j+1)/2)}.
\end{equation}
}
\begin{equation}
\begin{split}
V_0(Q)  &=\int\limits_Q\dv, \\
V_j(Q)&=\frac{1}{\omega_{j-1}\binko{d}{j}}
\int\limits_{\partial{Q}}\ds\gS_j\left(\kappa_1\ldots\kappa_{d-1}\right).
\end{split}
\label{eq:flatintegrals}
\end{equation} 
Here  $\dv$ and $\ds$  denote the  volume  element in $\EE^d$ and  the
surface element on $Q$, respectively, $\kappa_1$ to $\kappa_{d-1}$ are
the  boundary's $d-1$   principal curvatures,  and   $\gS_j$ is  the
$j$th  elementary symmetric  function   defined  by  the  polynomial
expansion
\begin{equation}
\prod_{i=1}^{d-1}(x+\kappa_i) =
\sum_{j=1}^{d}x^{d-j}
\gS_j\left(\kappa_1\ldots\kappa_{d-1}\right);
\end{equation}
hence $\gS_1=1$, $\gS_2=\kappa_1+\ldots+\kappa_{d-1}$, and so on up to
$\gS_d=\kappa_1\ldots\kappa_{d-1}$.          Table~\ref{tab:geometric}
summarizes  geometric interpretations of  the Minkowski functionals in
one, two and three dimensions.

\subsection{Spaces of constant curvature}

Let  us now consider the $d$--dimensional  space of constant curvature
$kK$. The  sign $k$ equals $+1$, $0$  or $-1$, for the spherical space
$\Ss^d$, the Euclidean space $\EE^d$ and the hyperbolic space $\HH^d$,
respectively.      $K$    is   a    positive   constant  of  dimension
$[\mbox{Length}]^{-2}$, hence its  inverse square root $K^{-1/2}$  can
be     interpreted     as         the    radius    of       curvature.
{}\scite{santalo:integralgeometry}  shows  how  to obtain  an integral
geometry      on   such  spaces.       Curvature   integrals   as   in
Equation~(\ref{eq:flatintegrals}) can  still  be defined, if   care is
taken to use the geodesic curvatures $\kappa_i$.  In the following, we
will call these quantities the Minkowski functionals in curved spaces.

However, some  of   the geometric  interpretations  are  altered  with
respect to the flat case.  While in flat  space the curvature integral
$V_d(Q)$ is equal to the Euler characteristic $\chi(Q)$, curved spaces
require a  generalized   Gauss--Bonnet Theorem  proved for   arbitrary
Riemannian    manifolds   by  {}\scite{allendoerfer:gaussbonnet}   and
{}\scite{chern:simple}.     The    theorem states   that     the Euler
characteristic is a linear combination of all Minkowski functionals as
defined by Equation~(\ref{eq:flatintegrals}),
\begin{equation}
\chi(Q) = \sum_{j=0}^d c_j V_j(Q),
\label{eq:gaussbonnet}
\end{equation}
with the coefficients $c_j$ given by
\begin{equation}
c_j = \left\{\begin{array}{cl}
\binko{d}{j}\dfrac{2(kK)^{(d-j)/2}}{\omega_{d-j}} &
\mbox{if $d-j$ even,}\\
0 & \mbox{if $d-j$ odd.}
\end{array}\right.
\end{equation}
Note that from the point of view of  Hadwiger's theorem, which is also
valid  on   curved  spaces,  all  linear   combinations   of Minkowski
functionals are equally  suitable as morphological descriptors, so one
may   both   use the   integrated    curvature  $V_d$ and   the  Euler
characteristic  $\chi$  as  the  last  Minkowski  functional.  In  the
following, we will consider  both  quantities, because the  integrated
geodesic   curvature    is  easier   to  calculate,    and   the Euler
characteristic  is easier to  interpret.   Obviously,  in the case  of
Euclidean space $\EE^d$, $k=0$  and all coefficients apart from  $c_d$
vanish,   so $\chi=V_d$  and the   original  Gauss--Bonnet theorem  is
recovered.

\subsection{Two--dimensional unit sphere}

We now focus attention  on the supporting space for  CMB sky maps, the
sphere  $\Ss^2$ of  radius  $R$.  The   parameters introduced  in  the
previous section now  take the values $d=2$  for the dimension, $k=+1$
for the curvature sign,  and $K=R^{-2}$ for  the absolute value of the
curvature.

Rewriting the  definition   in  Equation~(\ref{eq:flatintegrals}),  we
obtain the Minkowski functionals    for a set  $Q\subseteq\Ss^2$  with
smooth boundary $\partial{Q}$ by
\begin{multline}
V_0(Q)=\int\limits_Q\da,\\
V_1(Q)=\frac{1}{4}\int\limits_{\partial{Q}}\dl,\qquad 
V_2(Q)=\frac{1}{2\pi}\int\limits_{\partial{Q}}\dl\,\kappa,
\label{eq:lineintegrals}
\end{multline}
where $\da$ and  $\dl$ denote the surface  element of  $\Ss^2$ and the
line element  along  $\partial{Q}$,   respectively.  Being  a   linear
object,  the boundary $\partial{Q}$  has  only one geodesic  curvature
$\kappa$.

Using      the   generalized           Gauss--Bonnet    Theorem     in
Equation~(\ref{eq:gaussbonnet})   with   the  coefficients   for   two
dimensions substituted,  we can   calculate the Euler   characteristic
$\chi(Q)$ from the Minkowski functionals via
\begin{equation}
\chi(Q)=V_2(Q)+\frac{1}{2\pi{R^2}}V_0(Q).
\end{equation}
Note    that       by      inserting       the     definitions    from
Equation~(\ref{eq:lineintegrals}),      this    formula reproduces the
ordinary Gauss--Bonnet Theorem   for surfaces with a  smooth  boundary
embedded in three--dimensional flat space.

Let us now consider a  smooth  scalar field  $u(\bx)$ on $\Ss^2$,  for
example the temperature anisotropies of  the Microwave sky. We wish to
calculate the Minkowski functionals of  the excursion set $Q_\nu$ over
a given threshold $\nu$, defined by
\begin{equation}
Q_\nu=\left\{\,\bx\in\Ss^2\,|\,u(\bx)>\nu\,\right\}.
\end{equation}
The zeroth  Minkowski   functional  $V_0$,  i.e.\ the area,    can  be
evaluated by  integration of a Heaviside  step function over the whole
sphere
\begin{equation}
V_0(Q_\nu)=\int\limits_{\Ss^2}\da\,\Theta(u-\nu).
\end{equation}
The other Minkowski functionals are actually defined by line integrals
along the isodensity contour in Equation~(\ref{eq:lineintegrals}), but
they can  be transformed  to  surface integrals  by inserting  a delta
function, and the appropriate Jacobian.
\begin{equation}
\begin{split}
V_1(Q_\nu)
=\frac{1}{4}\int\limits_{\partial{Q_\nu}}\dl
&=\int\limits_{\Ss^2}\da\,\delta(u-\nu)|\grad{u}|\,\frac{1}{4},
\\
V_2(Q_\nu)
=\frac{1}{2\pi}\int\limits_{\partial{Q_\nu}}\dl\,\kappa
&=\int\limits_{\Ss^2}\da\,\delta(u-\nu)|\grad{u}|\,\frac{1}{2\pi}\kappa.
\end{split}
\end{equation}
Since the integrands can   now be written as  second--order invariants
(see Appendix~\ref{app:curvature} for   a detailed calculation of  the
geodesic curvature $\kappa$),  we   have succeeded in expressing   all
Minkowski functionals as surface   integrals  over the whole    sphere
$\Ss^2$,
\begin{equation}
V_j(Q_\nu) = \int\limits_{\Ss^2}\da\,\cI_j,
\label{eq:invariantintegrals}
\end{equation}
with integrands $\cI_j$ depending  solely on the threshold $\nu$,  the
field   value    $u$  and its   first--    and second--order covariant
derivatives.  In summary,
\begin{equation}
\begin{gathered}
\cI_0=\Theta(u-\nu), \\
\cI_1=\frac{1}{4}\delta(u-\nu)\sqrt{u_{;1}^2+u_{;2}^2}, \\
\cI_2=\frac{1}{2\pi}\delta(u-\nu)\frac{2u_{;1}u_{;2}u_{;12}-u_{;1}^2u_{;22}-u_{;2}^2u_{;11}}{u_{;1}^2+u_{;2}^2}.
\end{gathered}
\label{eq:invariants}
\end{equation}
In the  following, we will use  the surface densities of the Minkowski
functionals, that is divide by the area of $\Ss^2$.
\begin{equation}
v_j(\nu)
=\frac{1}{4\pi{R^2}} V_j(Q_\nu)
=\frac{1}{4\pi{R^2}}\int\limits_{\Ss^2}\da\,\cI_j.
\end{equation}

\subsection{Expectation values for a Gaussian random field}

Minkowski functionals  and other geometric characteristics of Gaussian
random fields are extensively studied by {}\scite{adler:randomfields}.
Analytical expressions for  the  average  Minkowski functionals  of  a
Gaussian   random   field in arbitrary    dimensions   were derived by
{}\scite{tomita:curvature}; in the special case of two dimensions, the
results for the isodensity contour at threshold $\nu$ are\footnote{The
function $\Phi(x)$ is the Gaussian  error function given by $\Phi(x) =
\frac{2}{\sqrt{\pi}} \int_0^x\dt\exp(-t^2)$.}
\begin{equation}
\begin{split}
v_0(\nu)&=\frac{1}{2}-\frac{1}{2}
\Phi\left(\frac{\nu-\mu}{\sqrt{2\sigma}}\right), \\
v_1(\nu)&=\frac{\tau^{1/2}}{8\sigma^{1/2}}
\exp\left(-\frac{(\nu-\mu)^2}{2\sigma}\right),
\\ v_2(\nu)&=\frac{\tau}{2\pi^{3/2}\sigma} \frac{\nu-\mu}{\sqrt{2\sigma}}
\exp\left(-\frac{(\nu-\mu)^2}{2\sigma}\right).
%
\end{split}
\label{eq:analytical}
\end{equation}

Note that  these  expressions contain   only three parameters,  namely
$\mu$, $\sigma$, and $\tau$.  All  three are  easily estimated from  a
given realization of the Gaussian  random field, by taking averages of
the field itself, its square, and the sum  of its squared derivatives;
then
\begin{equation}
\begin{split}
\mu    &=\left\langle{u}\right\rangle \\
\sigma &=\left\langle{u^2}\right\rangle-\mu^2 \\
\tau   &=\frac{1}{2}\left\langle{u_{;i}u_{;i}}\right\rangle.
\end{split}
\label{eq:parameterfrommap}
\end{equation}
With these relations and the spherical harmonics expansion of $u$, the
parameters  $\sigma$ and $\tau$ may  also  be calculated directly from
the angular power spectrum $C_\ell$, with the results
\begin{equation}
\begin{split}
\sigma &= \sum_{\ell=1}^\infty (2\ell+1)C_\ell, \\
\tau   &= \sum_{\ell=1}^\infty (2\ell+1)C_\ell\frac{\ell(\ell+1)}{2}.
\end{split}
\label{eq:parameterfrompower}
\end{equation}

\section{Estimating Minkowski functionals of pixelized CMB sky maps}
\label{sec:minkowski}

\begin{figure}
\bildchen{\linewidth}{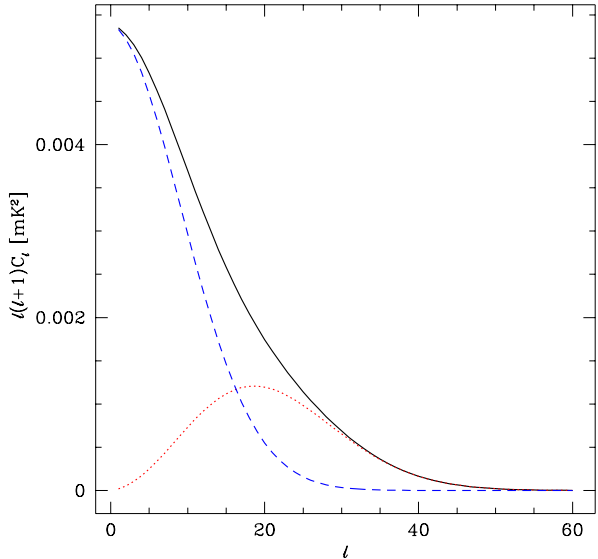}
\caption{
The angular power spectrum shown in this plot is used for all tests of
the  estimators presented in this section.   It was  chosen as a rough
model for  the DMR maps while remaining  analytically  tractable.  The
complete power spectrum, indicated by a solid  line, is the sum of two
contributions,  namely a  Harrison--Zeldovich spectrum  normalized  to
$C_{10}=(7\muK)^2$, convolved with a 7\deg\ FWHM beam to mimick cosmic
``signal'' (dashed  line), and random pixel  noise  (dotted line).  In
order  to   obtain a  regular  field and  suppress  noise the combined
spectrum  is  smoothed  with a   Gaussian filter; its   width for this
particular example is 3\deg.   Although the noise contribution has not
been completely removed, it   has  been considerably  reduced  without
affecting      the    signal       too         strongly.           See
Section~\protect\ref{sec:reduce}   for  a broader  discussion  of this
issue.
\label{fig:example.power} 
}
\end{figure}

Throughout this section,  we  will illustrate  the application  of our
method on a particular  random field.  In order  to stick to a simple,
analytically tractable model, we  generate  a Gaussian  random  field.
Its angular power spectrum $C_\ell$ is chosen to reproduce the salient
features of the DMR sky maps.  Hence, we start from angular components
given by the formula
\begin{equation}
C_{\ell,\rm powerlaw}=C_2\frac
{\Gamma\left(\ell+\frac{n-1}{2}\right)\Gamma\left(\frac{9-n}{2}\right)}
{\Gamma\left(\ell+\frac{5-n}{2}\right)\Gamma\left(\frac{n+3}{2}\right)},
\label{eq:powerlaw}
\end{equation}
derived     from  a   power--law   spectrum   $P(k)\propto{k^n}$    by
{}\scite{bond:statistics}, and smooth them with  a Gaussian filter  of
7\deg\  FWHM to  model  the DMR  beam.  White noise   with a fixed rms
fluctuation level of $3\muK$ is  then added to this ``cosmic signal'';
this is in    practice done on the    pixels in real  space, but   for
comparison we may also evaluate  the contribution to the angular power
spectrum, which is
\begin{equation}
C_{\ell,\rm noise}=(3\muK)^2
\end{equation}
independent of   $\ell$.  Finally,   a Gaussian smoothing    kernel of
variance $\omega^2$, given by
\begin{equation}
g_{\ell,\rm Gauss} 
= 
\exp\left(-\tfrac{1}{2}\omega^2\ell(\ell+1)\right)
\end{equation}
is applied to reduce  the noise level,  and to obtain a regular field.
Note       that     the   normalization        factor      $C_2$    in
Equation~(\ref{eq:powerlaw})  is   directly   related  to   the    CMB
quadrupole.   As   pointed  out   by  {}\scite{gorski:onresults}   its
particular  value  determined from the  COBE   DMR sky  maps is highly
dependent on the  spectral index $n$; therefore  it has become  common
usage    to  quote  the     multipole $C_{10}\approx(7\muK)^2$  as   a
sufficiently spectrum independent normalization.

So  the various contributions  to the  angular power spectrum $C_\ell$
for our example sum up to
\begin{equation}
C_\ell=g_{\ell,\rm Gauss}^2 \left(C_{\ell,\rm powerlaw} 
g_{\ell,\rm beam}^2 + C_{\ell,\rm noise}\right).
\end{equation}

In Figure~\ref{fig:example.power}, the contributions of ``signal'' and
noise are shown  separately, and combined to   the full spectrum,  all
after 3\deg\ smoothing .

\subsection{Estimators for a pixelized sky map}

In order to estimate the Minkowski  functionals from a discretized map
we  attempt     to     follow   the    prescription    outlined     by
{}\scite{schmalzing:beyond}  for   cubic  grids in  three--dimensional
Euclidean space.  Note that  their  first approach based  on Crofton's
formula  and  counting of   elementary cells  is  not  viable  since a
strictly regular pixelization of the  sphere does not exist.  However,
their second approach is based on  averaging over invariants analogous
to Equation~(\ref{eq:invariantintegrals}),  and  can easily be adapted
to  the sphere.   If the   random field is  sampled  at  $N$ pixels at
locations $\bx_i$ on the sphere,  we only need  to estimate the values
$\cI_j(\bx_i)$ of  the  invariants from Equation (\ref{eq:invariants})
at each location.

\subsubsection{Covariant derivatives}
\label{sec:covariant}

\def\t{\vartheta}
\def\p{\varphi}
\def\arg{(\t,\p)}
\def\alm{a_{\ell m}}
\def\Ylm{Y_{\ell m}}
\def\sumlm{\sum_{\ell=0}^\infty\sum_{m=-\ell}^\ell}

Using the  well--known   parametrization of the   unit  sphere through
azimuth angle $\t$ and  polar angle $\p$ we  can express the covariant
derivatives   at a  point   $\bx=(\t,\p)$  in  terms  of the   partial
derivatives\footnote{Note that we  use indices following  a semicolon,
such as $u_{;i}$  to  denote  covariant  differentiation of $u$   with
respect to the coordinate $i$, as opposed to partial derivatives where
we write indices following a comma, e.g.\ $u_{,i}$.};
\begin{equation}
\begin{split}
u_{;\t}&=                u_{,\t}, \\
u_{;\p}&=\frac{1}{\sin\t}u_{,\p}, \\
u_{;\t\t}&=                  u_{,\t\t}, \\
u_{;\t\p}&=\frac{1}{\sin\t}  u_{,\t\p}-\frac{\cos\t}{\sin^2\t}u_{,\p}, \\
u_{;\p\p}&=\frac{1}{\sin^2\t}u_{,\p\p}+\frac{\cos\t}{\sin\t}  u_{,\t}.
\end{split}
\end{equation}

The partial derivatives in turn are best calculated from the spherical
harmonics expansion
\begin{equation}
u\arg=\sumlm\alm\Ylm\arg.
\end{equation}
This is simply done by replacing the harmonic function $\Ylm$ with its
appropriate partial derivative.   Since the functions $\Ylm$ depend on
$\p$ via sine and cosine  functions only, the derivatives with respect
to   $\p$ can  be  obtained  analytically.   Partial derivatives  with
respect to  $\t$ are calculated  via recursion formulae constructed by
differentiating  the recursion for   the associated Legendre functions
$P_\ell^m$, given for example by {}\scite{abramowitz:handbook}.

\subsubsection{Integrals over invariants}

We still have to account for the finite number of sample points.  This
is done  by replacing the  delta function  with a  bin of finite width
$\Delta$,
\begin{equation}
\delta(u-\nu)\approx\frac{1}{\Delta}
\Eins_{[-\Delta/2,+\Delta/2]}(u-\nu),
\end{equation}
where $\Eins_A$  is the indicator   function of   the  set $A$,   with
$\Eins_A(x)=1$  for   $x\in{A}$, and $\Eins_A(x)=0$   otherwise.   The
integrals  summarized in Equation (\ref{eq:invariantintegrals})    are
then estimated by summation over all  pixels\footnote{We set the pixel
weight factors $w_i$ equal to 1,  but this may  be changed, if care is
taken to preserve $\sum_{i=1}^Nw_i=1$.}
\begin{equation}
v_j(\nu)\approx\frac{1}{N}\sum_{i=1}^Nw_i\cI_j(\bx_i).
\label{eq:integralbysum}
\end{equation}
For   incomplete sky  coverage  we must   restrict the  average to the
unmasked   pixels.    This   problem   is  adressed   in    detail  in
Section~\ref{sec:testing}.


\subsection{Testing the estimators}
\label{sec:testing}

\subsubsection{Complete sky coverage}

\begin{figure*}
\centerline{\bildchen{\bildchenwidth}{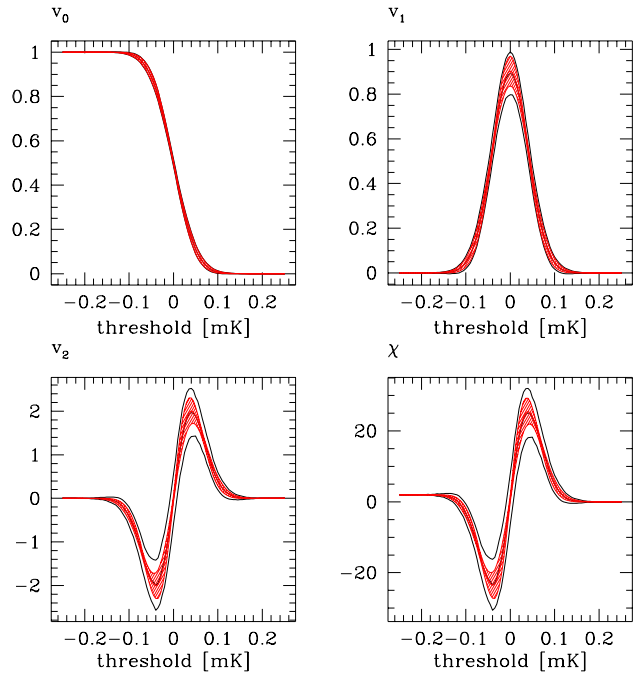}}
\caption{
Minkowski functionals for the  example  used throughout this  Section.
The areas  indicate  the  averages and $1\sigma$--fluctuations   of an
ensemble of 1,000 realizations computed on all 6,144 pixels of the DMR
sky   cube.       The     analytical  expectation       values      in
Equation~(\protect\ref{eq:analytical}) with parameters determined from
the           theoretical               power         spectrum     via
Equation~(\protect\ref{eq:parameterfrompower})  are   almost   exactly
reproduced by the mean values (central lines).  Fluctuations indicated
by the  shaded  area   are due  to  uncertainties  in   the  parameter
determination       from           a       single           map    via
Equation~(\protect\ref{eq:parameterfrommap}).    They account  for   a
large part of the overall fluctuations (empty area).
\label{fig:example.allsky}
}
\end{figure*}

To  begin with,  let us look   at the  example without simulating  the
restrictions         of         incomplete         sky       coverage.
Figure~\ref{fig:example.allsky}   shows      the    average  Minkowski
functionals of 1,000 realizations.  

Looking at the general features of all curves, it can be seen that the
area $v_0$ of hot spots decreases  monotonically from the value of one
at low threshold, when the whole sphere  belongs to the excursion set,
to a value of  zero at high thresholds which  are not passed by any of
the pixels.  The boundary length $v_1$ starts from a value of zero for
a completely filled sphere.    It  reaches a maximum   at intermediate
thresholds, where the excursion set forms an interconnected pattern of
patches and holes with  a very long boundary.   When the excursion set
becomes emptier and  emptier,  the boundary  length declines  back  to
zero.  For the  random  field  shown in  our  example,  the integrated
geodesic curvature $v_2$    behaves  largely  similar   to the   Euler
characteristic;  the  minor  differences  only  become appreciable for
fields with  fewer features.  Lastly,  the Euler characteristic $\chi$
at  low  thresholds has a  value  of two for  a  closed  sphere.  With
increasing threshold, the Euler   characteristic declines to  negative
values as   holes open   in  the  excursion set   and give  a negative
contribution.  This downward trend gradually  stops as individual  hot
spots  emerge, so  a minimum  develops, and  the Euler  characteristic
attains positive values.  Finally, more and  more hot spots fall below
the   growing threshold,   so  their  number    and   hence the  Euler
characteristic decreases again, reaching a final value of zero.

A  description of the  individual curves  can  be found in  the figure
caption.

\subsubsection{Uncertainties through incomplete sky coverage}

\label{sec:incomplete}

\begin{figure*}
\centerline{\bildchen{\bildchenwidth}{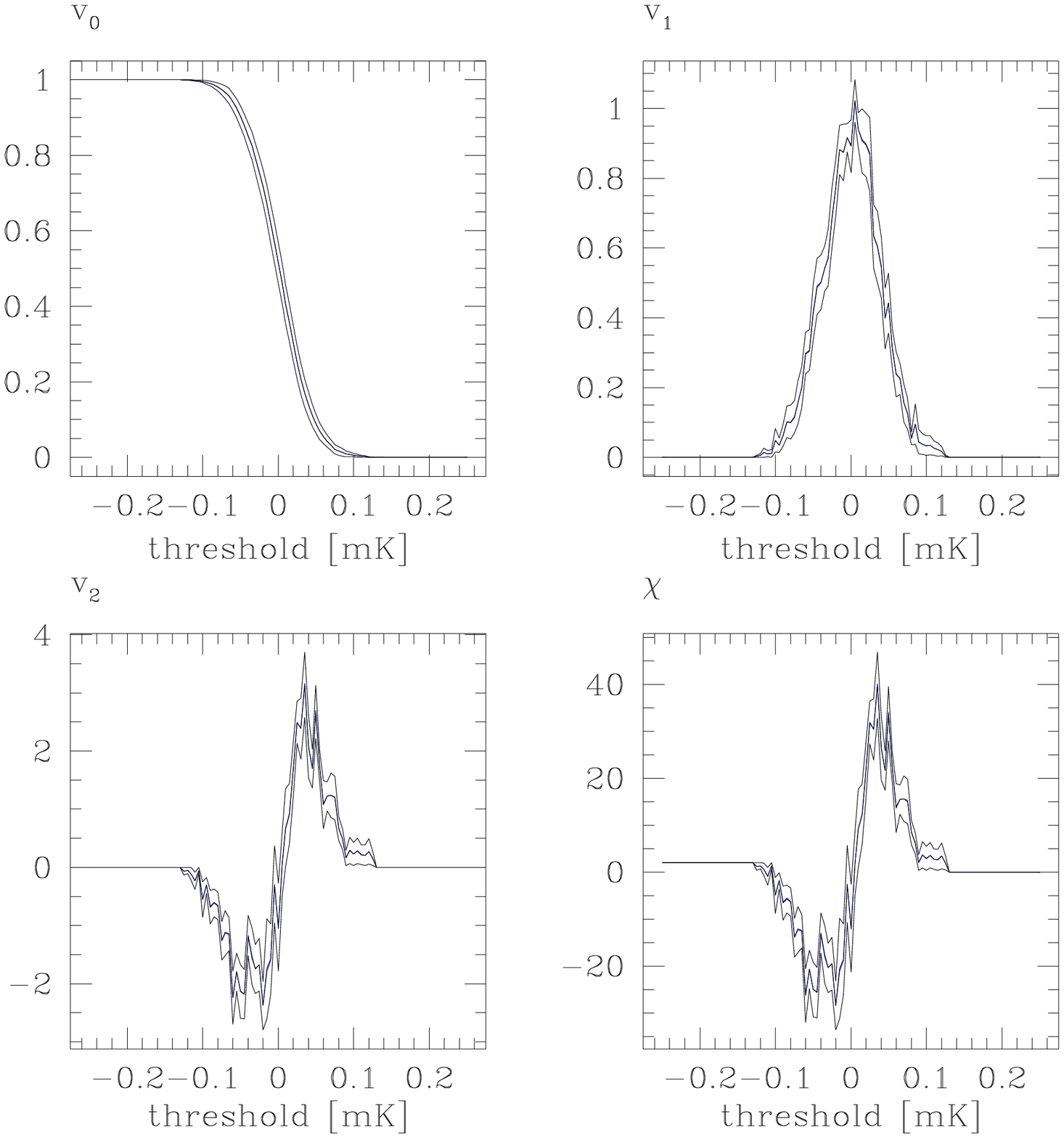}}
\caption{
This plot shows the uncertainties due to  a galactic cut.  In order to
separate them from   fluctuations between different  realizations in a
ensemble  (see Figure~\protect\ref{fig:example.allsky}),  we   chose a
single all--sky realization  (central line) of  the random  field, and
applied 1,000  straight  cuts  up to 30\deg\   latitude  with randomly
varying orientation  of the equatorial plane.   It turns out  that the
uncertainties caused by this deterioration of  statistics are at least
as strong as ensemble fluctuation.  However, the correct mean value is
still reproduced (the mean and the true value conincide in the central
line of the jagged area), so the estimator remains unbiased.
\label{fig:example.rotating}
}
\end{figure*}

In practice, a data set will suffer from incomplete  sky coverage.  In
order to estimate the uncertainties  introduced solely by the galactic
cut, we first contruct a single realization of the random field on the
whole   sky.  The  Minkowski functionals   for this  random field  are
calculated and   roughly   fit the   analytical   expectations,   with
fluctuations     consistent      with    the    areas    shown      in
Figure~\ref{fig:example.allsky}.  Then, we  apply a series of straight
galactic cuts  with  varying direction,  but  with constant  width  of
$30\deg$; this value reduces the number of  pixels to exactly half the
original value.  Figure~\ref{fig:example.rotating} shows a  comparison
of  the true values  for one field and  the fluctuations introduced by
the  sample variance  of the   rotating cuts.   Note  that the smaller
number of pixels does  increase the uncertainties,  but the average is
not affected -- the estimator remains unbiased.

\subsubsection{Boundary effects}
\label{sec:boundary}

\begin{figure*}
\centerline{\bildchen{\bildchenwidth}{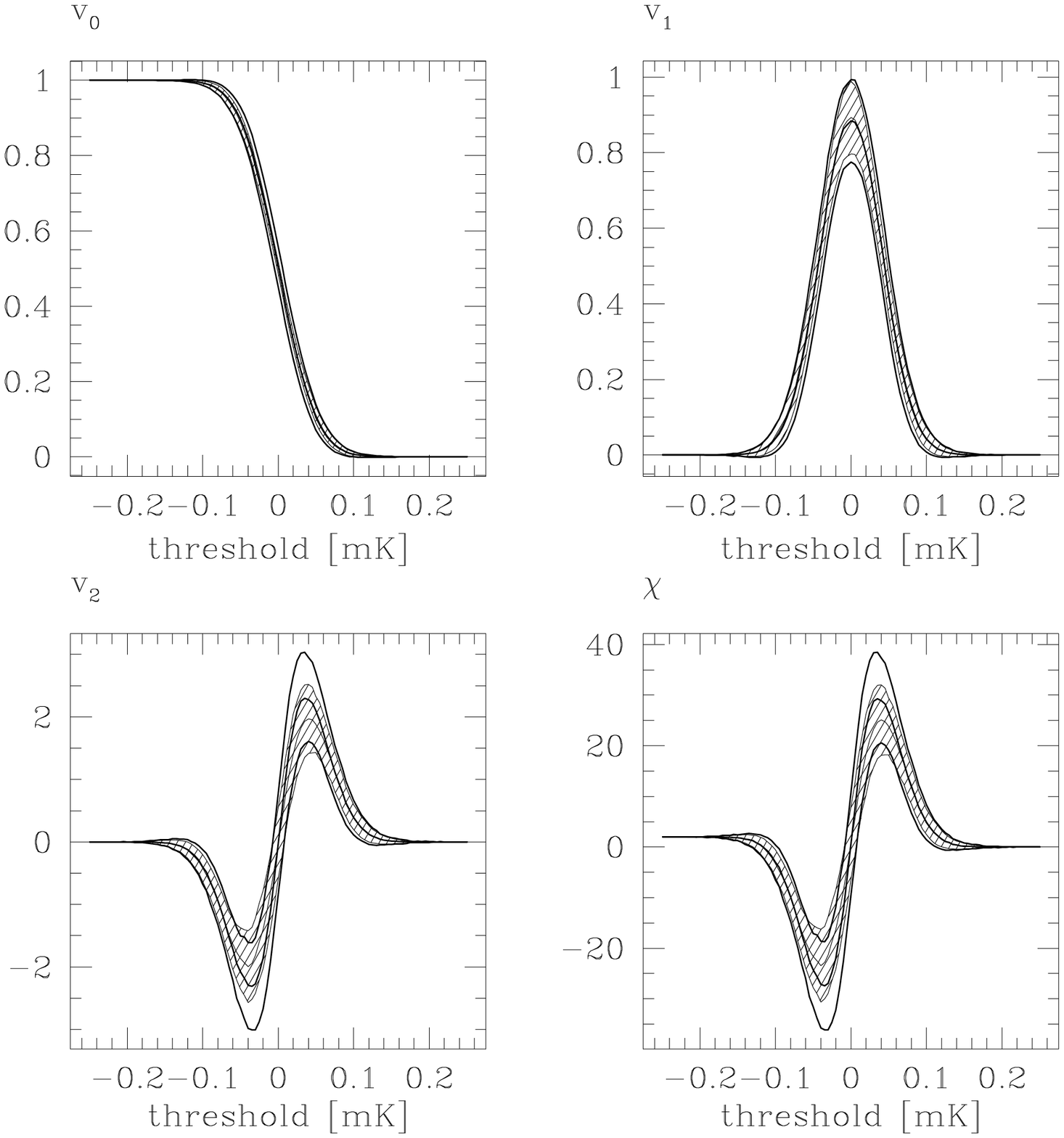}}
\caption{
If  the  galactic cut  is applied  {\em  before} smoothing  the random
field, as it should be  done for real data,  pixels in the vicinity of
the galactic  cut suffer from  severe contamination.  This  leads to a
visible bias in  the estimated Minkowski functionals, particularly the
integrated  curvature   $v_2$  and  the   Euler characteristic $\chi$,
although the area $v_0$ and the circumference $v_1$ are also affected.
The shaded area shows average  and  fluctuations for the all--sky  map
already presented in    Figure~\protect\ref{fig:example.allsky}, while
the empty  area with thicker   contour and central line  indicates the
same quantities for the maps biased through the cut.
\label{fig:example.galactic.bad}
}
\end{figure*}

\begin{figure*}
\centerline{\bildchen{\bildchenwidth}{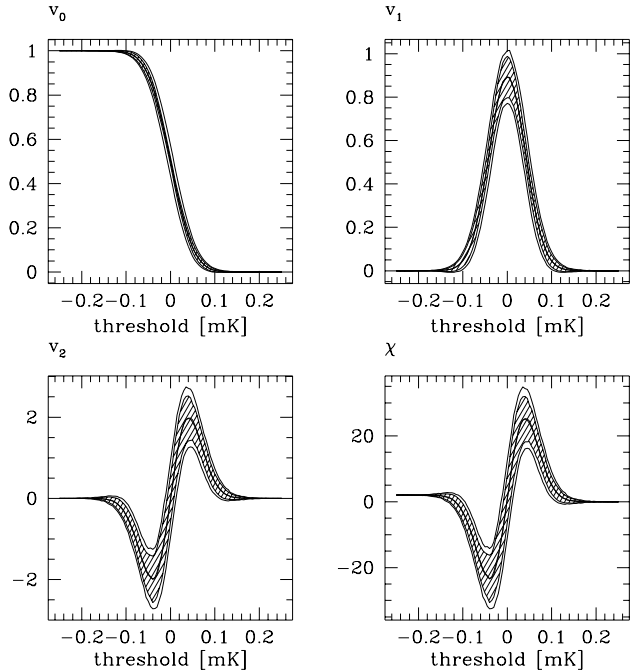}}
\caption{
The                biases               demonstrated                in
Figure~\protect\ref{fig:example.galactic.bad} can   be removed by also
smoothing the cut, and  excluding points where  the smoothed cut still
reaches a certain level.  In the example shown, this threshold was set
to a very restrictive 1\%; obviously the mean values agree completely.
In practice,  a  level as high as   5\% might still  produce  reliable
results.  Note that fluctuations have increased in comparison with the
results  obtained from  all--sky  maps,  as   already demonstrated  in
Figure~\protect\ref{fig:example.rotating}.
\label{fig:example.galactic}
}
\end{figure*}

\begin{figure*}
\centerline{\bildchen{\bildchenwidth}{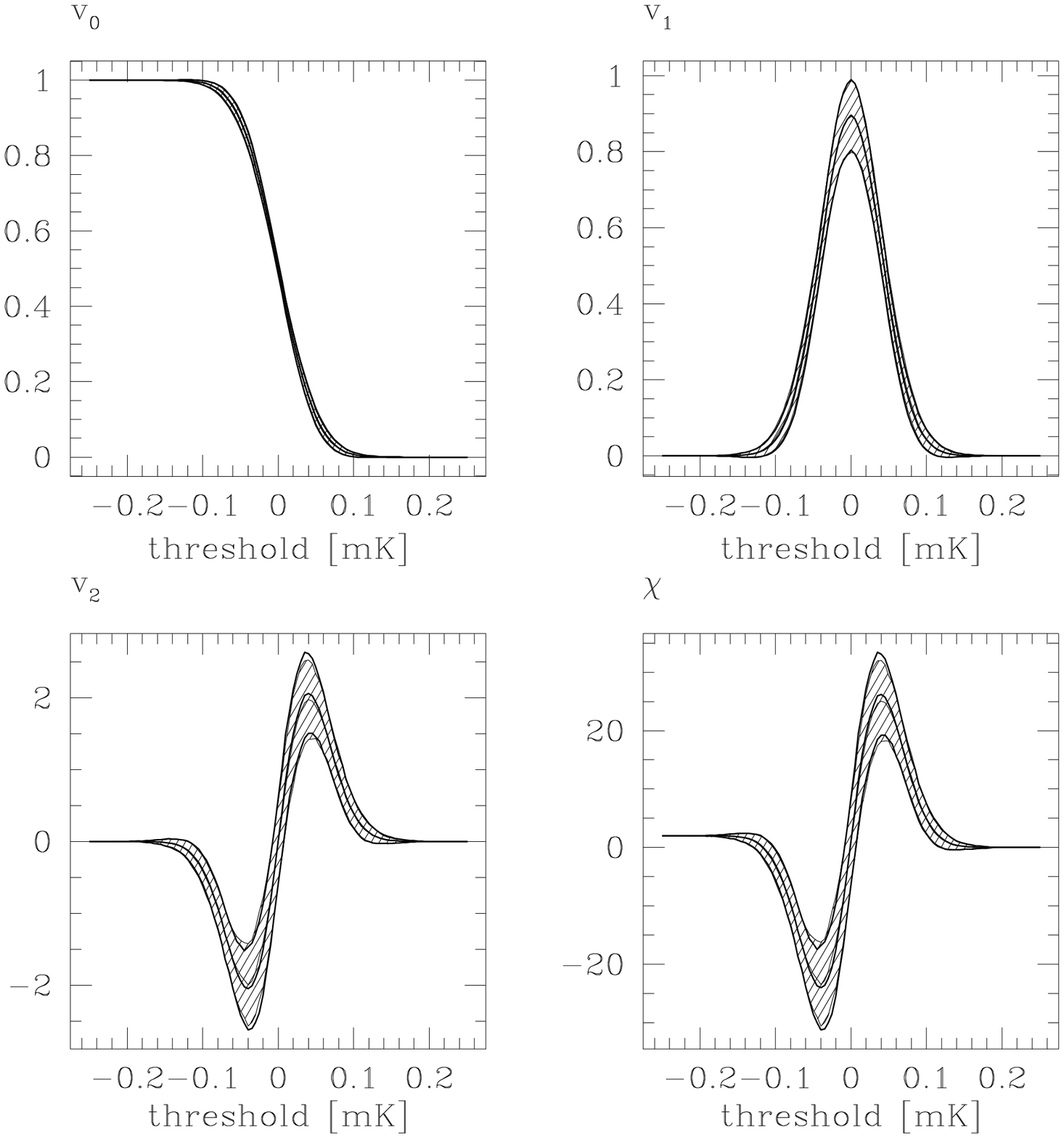}}
\caption{
The              same        quantities                as           in
Figure~\protect\ref{fig:example.galactic.bad}   are compared  in  this
plot, but with 200 random pixels excluded instead of the galactic cut.
Contamination  affects more pixels than in  the case of a galactic cut
--  if a residue of  1\% is allowed, 2,842  out of  the original 6,144
pixels  remain,  compared to 3,189 for   a galactic cut.  However, the
estimated Minkowski functionals are more robust.
\label{fig:example.point}
}
\end{figure*}

The previous   subsection  dealt with a random    field that was first
realized on the whole  sky, then smoothed with  a Gaussian filter, and
cut afterwards.   In  order  to determine  whether  the   galactic cut
affects the estimators  derived above, we  use the COBE DMR pixels and
the       customized     cut   from        the      four--year    data
{}\cite{bennett:fouryeardata}.  This time, we remove the pixels within
the galactic cut {\em before} the smoothing kernel is applied.

It  turns out that  by  using this   procedure, which is  actually the
correct one for mimicking real data, the galactic cut severely affects
the estimators,  and  leads  to  a systematic   bias  of  as  much  as
$1\sigma$.    Figure~\ref{fig:example.galactic.bad} shows the unbiased
results        from    all--sky      maps     already   displayed   in
Figure~\ref{fig:example.allsky},  in   comparison  the  biased  result
obtained with the na{\"\i}vely applied estimator.

A straightforward procedure to remove these biases from the estimators
is to further restrict the number of pixels to the ones that lie ``far
away'' from the cut.  In order to find them, we consider the indicator
function of the  cut itself, smooth  it with the  Gaussian filter, and
consider the values   at  pixels outside the  cut  as  their  level of
``contamination''.               Now   the    sums                from
Equation~(\ref{eq:integralbysum})  can be  restricted   to the  pixels
where  the   smoothed   cut     lies  below  a    certain   threshold.
Figure~\ref{fig:example.galactic} shows  the  results for  an  allowed
level  of 1\%; in practice, even  as much as 5\% produces sufficiently
unbiased estimates.  Note that while the  mean values agree completely
after applying the correction,  the   variance of the  estimators  has
increased,  simply because     fewer data points   result    in poorer
statistics.

Apart from the galactic  cut,  point source contamination  is  another
important     source       of     incomplete        sky      coverage.
Figure~\ref{fig:example.point}  shows the  bias introduced by omitting
200 randomly scattered  pixels.  Obviously,  the  effect is  much less
pronounced compared to the realizations excluding the galactic cut; in
fact the   differences between   the all--sky  realizations    and the
restricted realizations are barely visible.  Both the galactic cut and
the random point cut affect roughly 3,000 of the 6,144 DMR pixels with
a contribution of 1\% or above, so at  first sight our findings appear
inconsistent.  However, they  can   be explained with  the   prominent
geometric features -- namely, almost straight edges -- in the galactic
cut.  These are missing in a random  point distribution, so the errors
remain smaller and average out.

\section{Examples}
\label{sec:examples}

\subsection{The Earth}

\begin{figure}
\centerline{\bildchen{\linewidth}{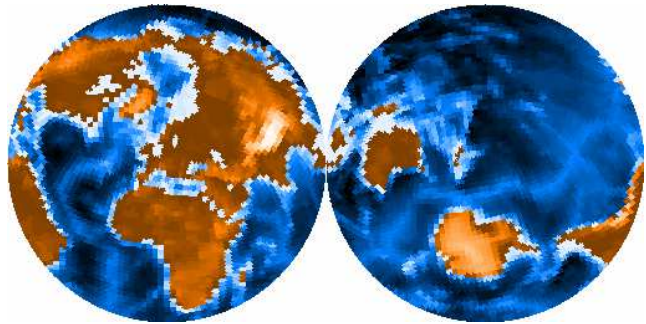}}
\caption{
This  map   shows the  earth's   topography  at  DMR  resolution.  The
corresponding     Minkowski  functionals       are     displayed    in
Figure~\protect\ref{fig:earth.minkowski}.
\label{fig:earth.map}
}
\end{figure}

\begin{figure*}
\centerline{\bildchen{\bildchenwidth}{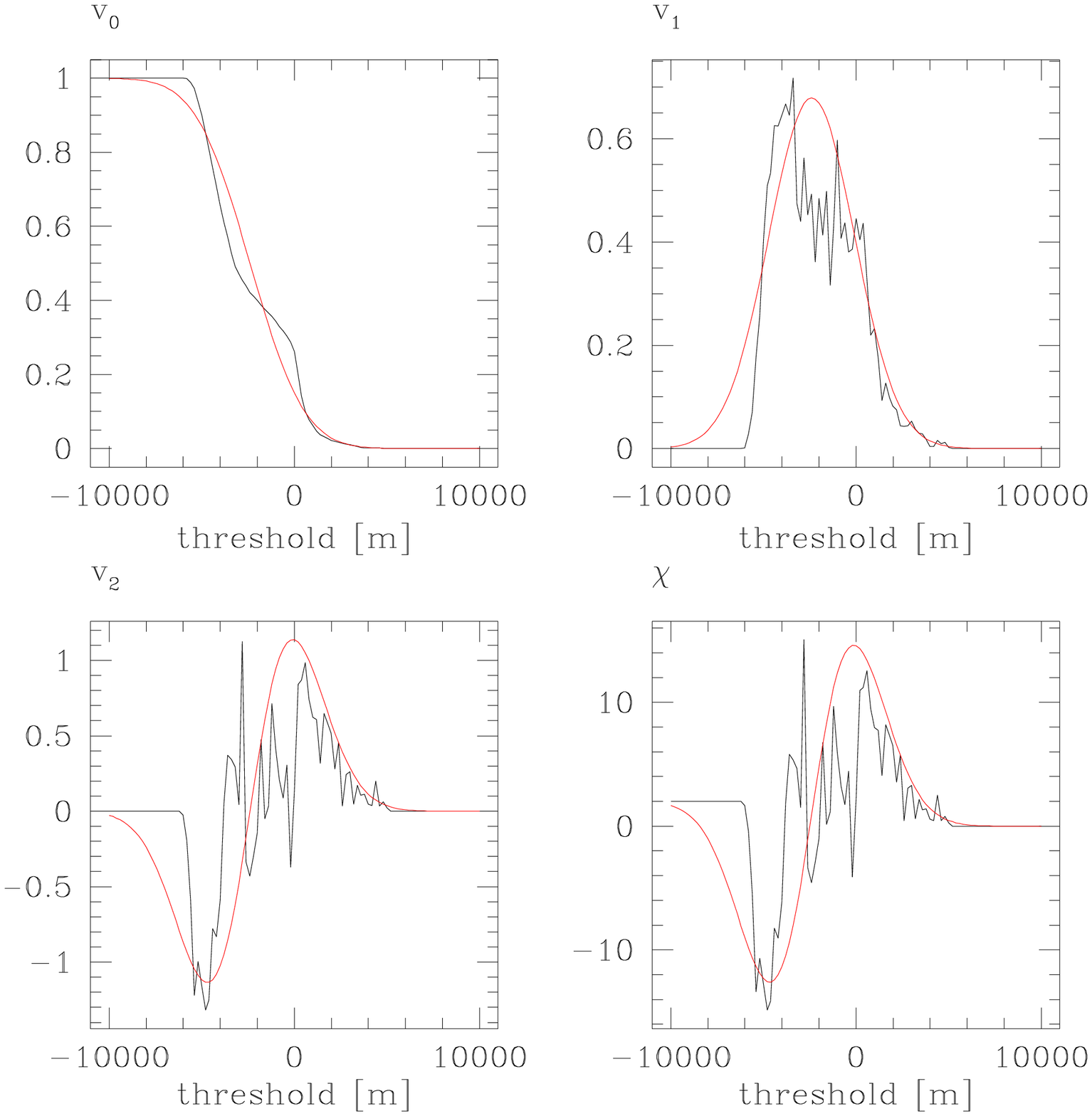}}
\caption{
Minkowski functionals of   the  earth's topography  from  the   map in
Figure~\protect\ref{fig:earth.map} with  a   1\deg\   Gaussian  filter
applied.  The  resulting  Minkowski  functionals (jagged  lines)  look
decidedly non--Gaussian;  compare   the analytical expectation  values
(smooth  lines).  Pronounced  features  in  the  Minkowski  functional
curves can be identified as corresponding to  the main features of the
earth's topography,  such as oceanic   ridges, continents and mountain
ranges.  See the main text for details.
\label{fig:earth.minkowski}
}
\end{figure*}

In order to provide a familiar example that does {\em not} look like a
Gaussian      random    field       even     at     DMR    resolution,
Figure~\ref{fig:earth.minkowski}  shows   the Minkowski functionals of
the earth's topography.  The  map (see Figure~\ref{fig:earth.map}) was
constructed  by binning  the Etopo5 data\footnote{The  Etopo5 database
gives elevations on  a cylindrical grid of 5   arcminute spacing.  The
data   files    may  be      obtained     from the  net     via   {\tt
ftp://walrus.wr.usgs.gov/pub/data/};  see also  {\em Data Announcement
88-MGG-02, Digital relief  of the Surface of the  Earth} by  the NOAA,
National Geophysical Data  Center, Boulder, Colorado,  1988.  }   into
the DMR pixels.

The  curves  reveal   a  number  of   characteristic  features.    All
functionals experience a fairly sharp change at a depth between 6,000m
and  5,000m, which is roughly  the average depth  of  the seafloor.  A
peak of several   1,000m width  and  almost  constant  height of   the
boundary   length $v_1$, and a    corresponding  minimum in the  Euler
characteristic  $\chi$ indicate the rise  of the oceanic ridges.  From
3,000m  below sea level to  slightly positive elevations, the boundary
length remains  largely constant, as  the continental shelfs rise from
the oceans; meanwhile,   the Euler characteristic fluctuates with  the
disappearance of the  oceanic  ridges, and the opening  of  shallower,
marginal parts of the oceans such as  the Mediterranean, the Carribean
sea or the Artic sea.    Most of the land  mass  does not rise  beyond
1,000m,   so all Minkowski  functionals gradually   decline after this
height; a   few small peaks  in  the  Euler  characteristic  may be --
cautiously --  identified with Antarctica,   the Rocky Mountains,  the
Andes, and the Himalaya.

\subsection{How smoothing leads to noise reduction}
\label{sec:reduce}

\begin{figure*}
\centerline{\bildchen{\bildchenwidth}{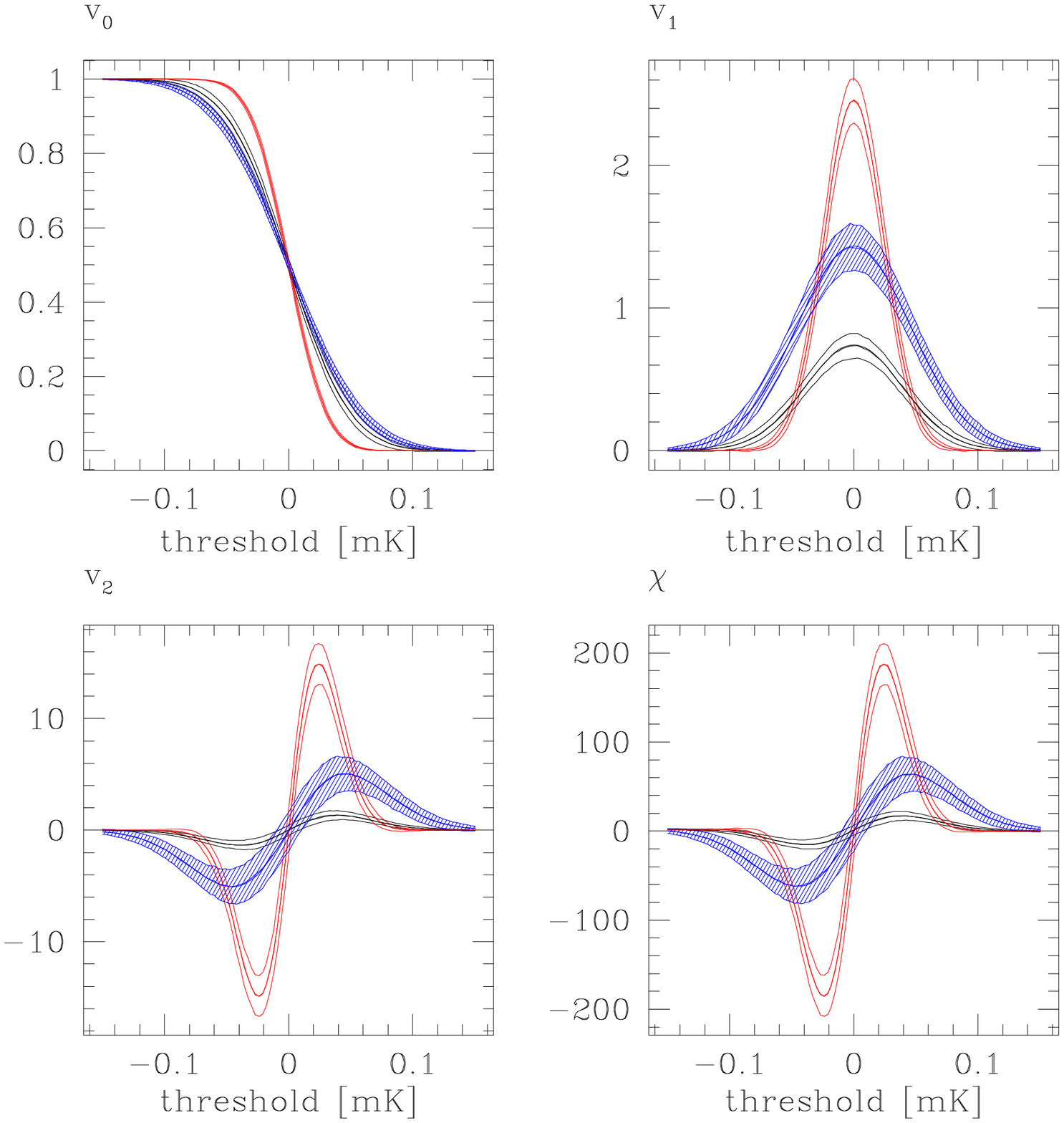}}
\caption{
Signal (shallow   curves, empty area),  noise (sharply  peaked curves,
empty area) and combination of both (shaded area) at 2\deg\ smoothing.
At this scale, the noise  contribution contains far more features than
the signal.  Moreover,   they   are strong   enough to persist    when
distributed over the whole range of the signal.  Hence the combination
of signal and noise   displays a completely different morphology  from
the pure signal contribution.
\label{fig:reduce.s2}
}
\end{figure*}

\begin{figure*}
\centerline{\bildchen{\bildchenwidth}{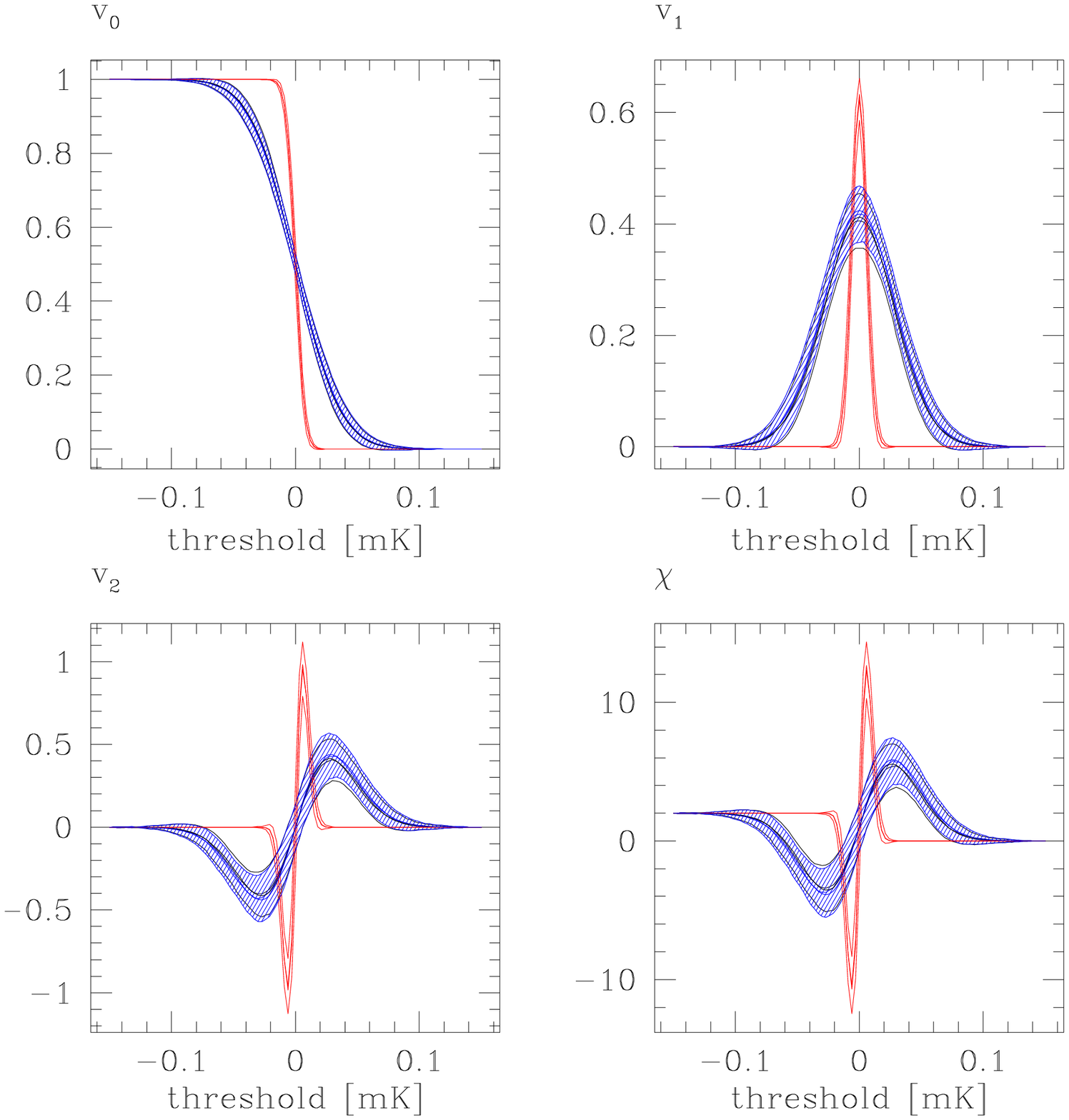}}
\caption{
Signal (shallow  curves, empty area),   noise (sharply peaked  curves,
empty area) and combination  of both (shaded area)  of the same random
field   as   in  Figure~\protect\ref{fig:reduce.s2}, but    at  8\deg\
smoothing.  Although   the pure noise map  still  features roughly two
dozen  extrema, their amplitude has been  reduced  to such extent that
the pure signal is hardly affected when combined with the noise.
\label{fig:reduce.s8}
}
\end{figure*}

In  order to obtain a  regular field, and to   reduce the level of the
additive  noise present  in  the  data, it   is necessary to  apply  a
smoothing  kernel   to the  data   before   calculating the  Minkowski
functionals.  Usually,  the choice  of a  particular  width is largely
arbitrary.     Here    we    show    the     example  introduced    in
Section~\ref{sec:examples} with different degrees of smoothing applied
to illustrate the behaviour  of Minkowski functionals in  the presence
of noise.

The situation  for  2\deg\  smoothing,  where  noise  still makes   an
appreciable contribution, is shown in Figure~\ref{fig:reduce.s2}.  The
surface area $v_0$ is  much  less affected  than the other   Minkowski
functionals;  this is due  to the  fact  that noise is  incoherent and
forms comparatively small hot and   cold spots.  However, these  spots
are almost as intense as the signal contribution, as  can be seen from
the almost equal  width of all  curves, and far  more numerous --  the
Euler characteristic for the noise  field alone  reaches a maximum  of
the order of 200.  Even though the extrema in  the pure noise maps are
spread  out  over  the whole range  of   thresholds when added  to the
signal,  and  hence their number at   a  specific threshold decreases,
their contribution   is  still sufficiently  high   to make the signal
appear completely different compared to the  combination of signal and
noise.

In Figure~\ref{fig:reduce.s8}, where the  results for 8\deg\ smoothing
are displayed, noise is  almost completely invisible in  comparison to
the signal.  Only about two dozen  extrema of either kind (compare the
extrema   of  the Euler characteristic)    remain, but since they have
become extremely shallow,   their contribution is  not significant any
more; the pure signal  and the combination of  signal and noise differ
only  marginally.    Unfortunately,  at a resolution   of   8\deg\ the
remaining signal does not carry too much cosmological information.

With   this example, the  behaviour   of  Minkowski functionals  under
filtering at different scales has only been hinted at.  The two filter
widths   of 2\deg\  and  8\deg\  are    chosen to   show  two extremal
possibilities, namely total dominance of  noise and total reduction of
noise.  In  practice, the  intermediate value of  3\deg\  turns out to
give   good  enhancement  of  signal,  while  preserving  small--scale
information as well.


\subsection{Analysis of the COBE DMR four--year data}

\begin{figure}
\centerline{
\bildchen{\linewidth}{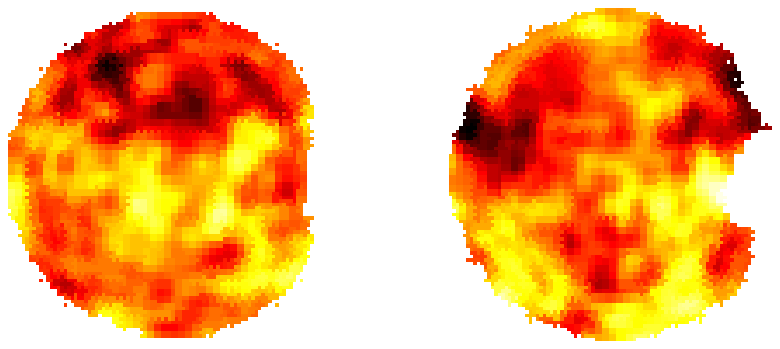}\hspace*{-\linewidth}%
\bildchen{\linewidth}{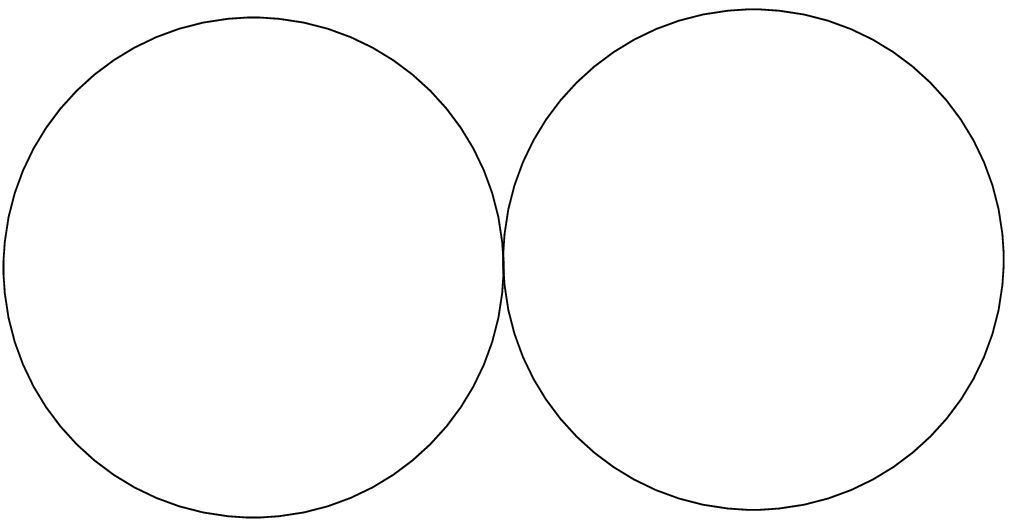}
}
\caption{
This panel shows the four--year data  from the COBE DMR 53GHz channel,
with a customized galactic cut and a smoothing  filter of 3\deg\ width
applied.     Figure~\protect\ref{fig:dmr.minkowski}   displays     the
corresponding Minkowski functionals calculated both from this map, and
from the all--sky map with galactic signal dominating.
\label{fig:dmr.map}
}
\end{figure}

\begin{figure*}
\centerline{\bildchen{\bildchenwidth}{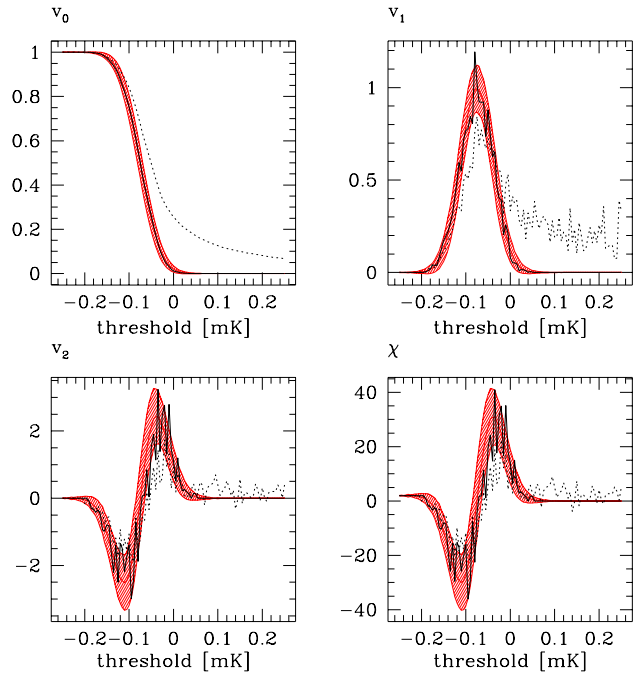}}
\caption{
Minkowski  functionals of the  four--year  COBE DMR  map in  the 53GHz
channel.  As expected, the full  map of 6,144  pixels (dashed line) is
severely affected by galactic emission; this results in long tails for
all  Minkowski  functionals.  Using  a  galactic cut,  the functionals
determined from  the   map  (solid  line) become   consistent   with a
stationary Gaussian random field.  The shaded  area indicates the mean
and variance  of 1,000 realizations  of a  Gaussian  random field with
Harrison--Zel'dovich   spectrum and  pixel noise; normalizations  were
chosen to  reproduce   the  parameters   $\mu$, $\sigma$   and  $\tau$
determined from the data via Equation~(\ref{eq:parameterfrommap}).
\label{fig:dmr.minkowski}
}
\end{figure*}

\begin{figure}
\centerline{\bildchen{\linewidth}{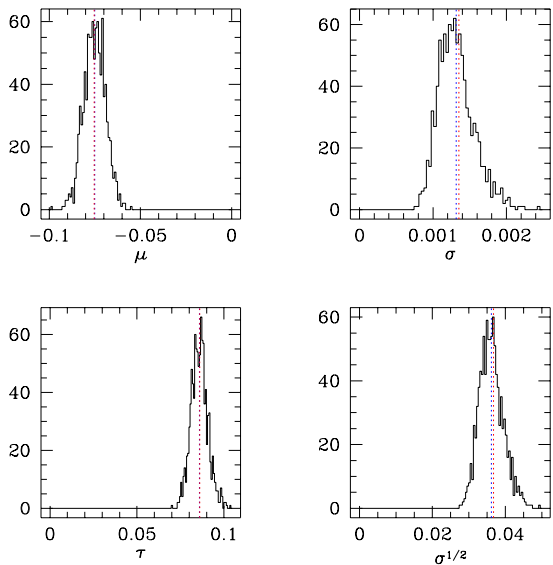}}
\caption{
Uncertainties in the  parameter estimates.  Parameters were determined
from each of the 1,000 simulated DMR maps, and binned into histograms.
In  order to give an  idea of the amount  of uncertainty, the range is
extended to  include zero in all   plots.  Obviously, the fluctuations
are still fairly large in comparison to the mean value; to be precise,
the relative rms fluctuation is roughly 8\% for  $\mu$, almost 6\% for
$\tau$, and more than 9\% for $\sqrt{\sigma}$.  
\label{fig:dmr.parameter} \label{lastpage}
}
\end{figure}

As a last example, let us take  a look at data that  are both real and
cosmologically relevant.  Figure~\ref{fig:dmr.map} shows  a map of the
microwave sky as seen at 53GHz by the COBE  satellite after four years
of observing {}\cite{bennett:fouryeardata}.   The data  are restricted
to 3,189 pixels receiving less than  1\% from a smoothed galactic cut,
when   a   Gaussian     filter  of    3\deg\   width   is     applied.
Figure~\ref{fig:dmr.minkowski} displays the corresponding    Minkowski
functionals.  Obviously, the analysis carried  out on all 6,144 pixels
is dominated  by galactic emission, while  the field with the galactic
cut applied is consistent with the assumption of a stationary Gaussian
random field.    However,    this  result  should   be  considered   a
illustration of the method rather than  conclusive evidence, since our
brief analysis probes a scale of roughly 4\deg\  (given by the squared
sum of 2.6\deg\ pixel size and 3\deg\ Gaussian filter width).

As stated above, considerable uncertainties are introduced through the
estimates of the parameters  $\mu$,  $\sigma$ and $\tau$ entering  the
analytical  expectation values  for   the Minkowski functionals of   a
Gaussian  random  field.   In  order  to  make  this   statement  more
quantitative, Figure~\ref{fig:dmr.parameter} summarizes the parameters
determined from  the 1,000 mock  realizations used for the shaded area
in Figure~\ref{fig:dmr.minkowski}.  Relative  errors for the  relevant
parameters lie in the range of five to ten  per cent, which is not too
bad  considering that  little more than  3,000 data  points enter  our
analysis.

\section{Summary and outlook}
\label{sec:outlook}

We have introduced Minkowski functionals of isotemperature contours as
a novel tool    to characterize the  morphology of   Cosmic  Microwave
Background sky maps.

Using  the framework of  integral geometry in  curved  spaces, we were
able to   clarify  the geometric   interpretations   of all  Minkowski
functionals in two dimensions.   Writing all Minkowski  functionals as
spatial averages  over  invariants formed from  covariant derivatives,
lead to   simple formulae for  estimators  applicable to pixelized sky
maps,  including   a straightforward  prescription   for  dealing with
incomplete  sky    coverage.  Finally,  analytical   formulae  for all
Minkowski functionals of the excursion sets of a Gaussian random field
could be provided.

The study of   a simplified yet  realistic model  served  to test  the
theoretically derived estimators in their application to simulated CMB
sky maps.  Among other tests, we checked whether the estimators remain
unbiased when incomplete data is smoothed over the  edge of a galactic
cut,   and  found  a prescription  to    deal with this  problem while
preserving as much information as possible.

A number of examples provided further illustrations of the application
of our  method, and showed how to  interpret  the calculated Minkowski
functionals in close touch with  the analyzed data sets.  The analysis
of the earth's topography using Minkowski functionals explained how to
cast the bridge from the behaviour of  the Minkowski functional curves
to  outstanding  features in the  underlying random  field.  As a more
serious application, we  showed   the successive reduction   of  noise
through Gaussian smoothing with increasing filter width -- in the end,
a complete removal of the noise effects from the Minkowski functionals
is obtained.  The final example briefly analyzed  a COBE DMR map, with
the not particularly surprising   result that the field  is consistent
with a Gaussian random field on degree scales.

Minkowski functionals  combine  the benefits  of a sound  mathematical
framework and well--understood analytical possibilities with intuitive
interpretations and   easy  applicability to  real  data.  Hence  they
qualify   as a  method  suited to  study the  Microwave  sky at higher
resolution, where  the obstacle of  poor  statistics should  not be an
issue.    While experiments to   obtain high--resolution maps of large
regions    of    the     sky    are     still   under      development
({}\pcite{bersanelli:cobrassamba}, {}\pcite{bennett:map}), testing the
Minkowski functionals  on  simulations is  an  important task for  the
future.    In particular, we need    to assess their  power to  detect
non--Gaussianity, and find  the possibilities to estimate cosmological
parameters from Minkowski functionals  in an approach complementary to
the power spectrum analysis.

\section*{Acknowledgements}

It is a pleasure to  thank Thomas Buchert, Martin Kerscher,  R\"udiger
Kneissl and Herbert Wagner  for enlightening discussions, comments and
suggestions.   We are indebted towards Peter  Coles for inspiring this
collaboration.  JS acknowledges    hospitality by TAC  during a  visit
where  part   of this work   was  prepared.   The  COBE  datasets were
developed by the  NASA Goddard Space  Flight Center under the guidance
of the COBE Science Working Group and were provided by the NSSDC.

\providecommand{\bysame}{\leavevmode\hbox to3em{\hrulefill}\thinspace}
\providecommand{\aanda}{Astron.\  Astrophys.}
\providecommand{\afz}{Astrophysics}
\providecommand{\apjl}{Ap.\ J.\ Lett.}
\providecommand{\apj}{Ap.\ J.}
\providecommand{\baas}{Bull.\ American Astron.\ Soc.}
\providecommand{\mnras}{Mon.\ Not.\ Roy.\ Astron.\ Soc.}

\appendix

\section{Geodesic curvature of an isodensity contour}
\label{app:curvature}

Consider  a   scalar field  $u$ on   a two--dimensional differentiable
manifold  $\gM$.  We wish  to calculate the  geodesic curvature of the
isodensity contour passing through a  point $\bx_0\in\gM$. To do this,
we use a procedure outlined  by {}\scite{koenderink:scalespace}.  In a
sufficiently small    neighbourhood  we can  always  find  an explicit
parametrization  $\bx(t)$    of   the   isodensity    contour,    with
$\bx(t=0)=\bx_0$.  The corresponding  threshold is $\nu=u(\bx_0)$, and
therefore the contour is implicitly described by
\begin{equation}
u(\bx(t))=\nu.
\label{eq:iso}
\end{equation}
It follows by covariant differentiation with  respect to the parameter
$t$  that\footnote{The overdot $\dot{}$  denotes differentiation  with
respect  to the parameter $t$,  and summation over pairwise indices is
understood.}
\begin{equation}
u_{;i}\dot{x}_i=0
\end{equation}
must  hold.    So  we    can   choose  the   tangent  vector\footnote{
$\epsilon_{ij}$ is    the  totally antisymmetric   second--rank tensor
normalized to $\epsilon_{12}$=1.}
\begin{equation}
\dot{x}_i=\epsilon_{ij}u_{;j}.
\end{equation}
Actually this   choice  is not   unique and  reflects  the freedom  of
parametrization; however, care  must be taken   to orient the  tangent
vector   towards  regions of  lower   values  of $u$.  Differentiating
Equation~(\ref{eq:iso}) a second time we obtain
\begin{equation}
u_{;ij}\dot{x}_i\dot{x}_j+u_{;k}\ddot{x}_k=0
\end{equation}
whence we can  now evaluate  the geodesic  curvature $\kappa$ of   the
isodensity contour via the well--known formula
\begin{equation}
\kappa=\frac{\dot{x}_i\epsilon_{ij}\ddot{x}_j}{(\dot{x}_k\dot{x}_k)^{3/2}}.
\end{equation}
As our final result we obtain
\begin{equation}
\kappa
=\frac{2u_{;1}u_{;2}u_{;12}-u_{;1}^2u_{;22}-u_{;2}^2u_{;11}}
{(u_{;1}^2+u_{;2}^2)^{3/2}}
.
\end{equation}
Note that this formula contains the covariant derivatives of $u$ and
therefore holds for any manifold $\gM$, regardless of the metric.

\section{Average Minkowski functionals for a Gaussian random field}
\label{app:average}

Isodensity contours of a  Gaussian random field have  been excessively
studied ever  since the  works of {}\scite{doroshkevich:gauss}  on the
genus.   Comprehensive   overviews   can  be  found  in   the  book by
{}\scite{adler:randomfields}    or     the      famous  BBKS     paper
{}\cite{bardeen:gauss}.    A  highly  instructive  derivation  of  the
average values  for all Minkowski  functionals  in arbitrary dimension
can  be found  in  {}\cite{tomita:statistics}.  Nevertheless, we  will
outline a   calculation  directly  related  to  our  approach   to the
numerical evaluation.

A    homogeneous  Gaussian random   field  $u$   with zero   mean on a
two--dimensional  manifold   $\gM$ is fully  described  by correlation
function   $\xi(r)$.  We   wish  to calculate   the average  Minkowski
functionals $v_j(\nu)$ of  an isodensity  contour  to the  threshold
$\nu$.   Because  of    Equations~(\ref{eq:invariantintegrals})    and
(\ref{eq:invariants}) it is  sufficient to know the  joint probability
distribution of  the field's  value  itself and  the derivatives up to
second order at some fixed point.

According   to  {}\scite{adler:randomfields}  these  six variables are
jointly Gaussian   distributed.  Hence their  probability distribution
function can be written   in concise form  by  arranging them into   a
vector  $\bu=\left(u,u_{;1},u_{;2},u_{;11},u_{;22},u_{;12}\right)$; we
have
\begin{equation}
P(\bu)=
\frac{1}{\sqrt{(2\pi)^d\det\bsigma}}
\exp\left(-\frac{1}{2}\bu^T\bsigma^{-1}\bu\right),
\end{equation}
with       the    covariance      matrix     $\bsigma$    taken   from
{}\scite{tomita:statistics}
\begin{equation}
\bsigma=\left(\begin{array}{cccccc}
\sigma & 0 & 0 & -\tau & -\tau & 0 \\
0 & \tau & 0 & 0 & 0 & 0 \\
0 & 0 & \tau & 0 & 0 & 0 \\
-\tau & 0 & 0 & \ups & \ups/3 & 0 \\
-\tau & 0 & 0 & \ups/3 & \ups & 0 \\
0 & 0 & 0 & 0 & 0 & \ups/3
\end{array}\right).
\end{equation}
Note  that  the  parameters   $\sigma=\xi(0)$,  $\tau=|\xi''(0)|$  and
$\ups=\xi''''(0)$ all depend on the correlation function $\xi(r)$.

Now we can perform the averages
\begin{equation}
\begin{split}
v_0(\nu)&=\int\rd\bu P(\bu)\Theta(u-\nu) \\
v_1(\nu)&=\int\rd\bu P(\bu)\delta(u-\nu)\sqrt{u_{;1}^2+u_{;2}^2} \\
v_2(\nu)&=\int\rd\bu P(\bu)\delta(u-\nu)\frac{2u_{;1}u_{;2}u_{;12}-u_{;1}^2u_{;22}-u_{;2}^2u_{;11}}{u_{;1}^2+u_{;2}^2}
\end{split}
\end{equation}
by  straightforward integration, and   have  recovered the results  of
{}\scite{tomita:statistics} in two dimensions
\begin{equation}
\begin{split}
v_0(\nu)&=
\tfrac{1}{2}-\tfrac{1}{2}\Phi\left(\frac{\nu}{\sqrt{2\sigma}}\right), \\
v_1(\nu)&= \tfrac{\pi}{4}\frac{\lambda}{\sqrt{2\pi}}
\exp\left(-\frac{\nu^2}{2\sigma}\right), \\
v_2(\nu)&=\frac{\lambda^2\nu}{\sqrt{2\pi\sigma}}
\exp\left(-\frac{\nu^2}{2\sigma}\right).
\end{split}
\end{equation}
As stated in the main text, the result depends on only two parameters,
namely
\begin{equation}
\sigma=\xi(0),\qquad
\lambda=\sqrt{\dfrac{|\xi''(0)|}{2\pi\xi(0)}}.
\end{equation}

\bsp

\label{lastpage}


\begin{thebibliography}{{ter Haar~Romeny \bgroup et al.\egroup }{1991}}

\bibitem[\protect\citefmt{Abramowitz \& Stegun}{1970}]{abramowitz:handbook}
Abramowitz, M.~X. \& Stegun, I.~A., \emph{Handbook of mathematical functions},
  9th ed., Dover Publications, New York, 1970.

\bibitem[\protect\citefmt{Adler}{1981}]{adler:randomfields}
Adler, R.~J., \emph{The geometry of random fields}, John Wiley \& Sons,
  Chichester, 1981.

\bibitem[\protect\citefmt{Allendoerfer \&
  Weil}{1943}]{allendoerfer:gaussbonnet}
Allendoerfer, C.~B. \& Weil, A., Trans.\ Amer.\ Math.\ Soc. \textbf{53} (1943),
  101--129.

\bibitem[\protect\citefmt{Bardeen \bgroup et al.\egroup }{1986}]{bardeen:gauss}
Bardeen, J.~M., Bond, J.~R., Kaiser, N., \& Szalay, A.~S.,
  {\apj} \textbf{304} (1986), 15--61.

\bibitem[\protect\citefmt{Bennett \bgroup et al.\egroup }{1995}]{bennett:map}
Bennett, C.~L., Hinshaw, G., Jarosik, N.~C., Mather, J., Meyer, S.~S., Page,
  L., Skillman, D., Spergel, D.~N., Wilkinson, D.~T., \& Wright, E.~L.,
  {\baas} \textbf{187}
  (1995), no.~71.09, 1385.

\bibitem[\protect\citefmt{Bennett \bgroup et al.\egroup
  }{1996}]{bennett:fouryeardata}
Bennett, C.~L., Banday, A., G{\'o}rski, K.~M., Jackson, P.~D., Keegstra, P.~B.,
  Kogut, A., Smoot, G.~F., Wilkinson, D.~T., \& Wright, E.~L.,
  {\apjl} \textbf{464} (1996), L1--L4.

\bibitem[\protect\citefmt{Bersanelli \bgroup et al.\egroup
  }{1996}]{bersanelli:cobrassamba}
Bersanelli, M., Bouchet, F.~R., Efstathiou, G., Griffin, M., Lamarre, J.~M.,
  Mandolesi, N., Norgaard-Nielsen, H.~U., Pace, O., Polny, J., Puget, J.~L.,
  Tauber, J., Vittorio, N., \& Volont{\'e}, S., \emph{{COBRAS/SAMBA}. {A}
  mission dedicated to imaging the anisotropies of the cosmic microwave
  background. {R}eport on the phase {A} study}, European Space Agency, February
  1996.

\bibitem[\protect\citefmt{Blaschke}{1936}]{blaschke:I}
Blaschke, W., \emph{{I}ntegralgeometrie. {E}rstes {H}eft}, Bernd G.\ Teubner,
  Leipzig, Berlin, 1936.

\bibitem[\protect\citefmt{Bond \& Efstathiou}{1987}]{bond:statistics}
Bond, J.~R. \& Efstathiou, G., {\mnras} \textbf{226} (1987), 655--687.

\bibitem[\protect\citefmt{Chern}{1944}]{chern:simple}
Chern, S.-S., Ann.\ of Math. \textbf{45} (1944), 747--752.

\bibitem[\protect\citefmt{Coles \& Barrow}{1987}]{coles:nongaussian}
Coles, P. \& Barrow, J.~D., {\mnras} \textbf{228} (1987), 407--426.

\bibitem[\protect\citefmt{Coles}{1988}]{coles:statistical}
Coles, P., {\mnras}
  \textbf{234} (1988), 509--531.

\bibitem[\protect\citefmt{Colley \bgroup et al.\egroup
  }{1996}]{colley:topology}
Colley, W.~N., {Gott III}, J.~R., \& Park, C., {\mnras} \textbf{281} (1996), L82--L84.

\bibitem[\protect\citefmt{Doroshkevich}{1970}]{doroshkevich:gauss}
Doroshkevich, A.~G., {\afz} \textbf{6}
  (1970), 320--330.

\bibitem[\protect\citefmt{G{\'o}rski \bgroup et al.\egroup
  }{1994}]{gorski:onresults}
G{\'o}rski, K.~M., Hinshaw, G., Banday, A.~J., Bennett, C.~L., Wright, E.~L.,
  Kogut, A., Smoot, G.~F., \& Lubin, P.~M., {\apjl} \textbf{430} (1994), L89--L92.

\bibitem[\protect\citefmt{G{\'o}rski \bgroup et al.\egroup
  }{1996}]{gorski:power}
G{\'o}rski, K.~M., Banday, A.~J., Bennett, C.~L., Hinshaw, G., Kogut, A.,
  Smoot, G.~F., \& Wright, E.~L., {\apjl}
  \textbf{464} (1996), L11--L16.

\bibitem[\protect\citefmt{{Gott III} \bgroup et al.\egroup
  }{1990}]{gott:topologymicrowave}
{Gott III}, J.~R., Park, C., Juszkiewicz, R., Bies, W.~E., Bennett, D.~P.,
  Bouchet, F.~R., \& Stebbins, A., {\apj}
  \textbf{352} (1990), 1--14.

\bibitem[\protect\citefmt{Hadwiger}{1957}]{hadwiger:vorlesung}
Hadwiger, H., \emph{Vorlesungen {\"u}ber {I}nhalt, {O}berfl{\"a}che und
  {I}soperimetrie}, Springer Verlag, Berlin, 1957.

\bibitem[\protect\citefmt{Hinshaw \bgroup et al.\egroup
  }{1995}]{hinshaw:threepoint}
Hinshaw, G., Banday, A.~J., Bennett, C.~L., G{\'o}rski, K.~M., \& Kogut, A.,
  {\apjl} \textbf{446} (1995), L67--L70.

\bibitem[\protect\citefmt{Hinshaw \bgroup et al.\egroup
  }{1996}]{hinshaw:twopoint}
Hinshaw, G., Banday, A.~J., Bennett, C.~L., G{\'o}rski, K.~M., Kogut, A.,
  Lineweaver, C.~H., Smoot, G.~F., \& Wright, E.~L.,
  {\apjl} \textbf{464} (1996), L25--L28.

\bibitem[\protect\citefmt{Kerscher \bgroup et al.\egroup
  }{1997a}]{kerscher:fluctuations}
Kerscher, M., Schmalzing, J., Buchert, T., \& Wagner, H.,
  {\aanda} (1997), accepted,
  astro-ph/9704028.

\bibitem[\protect\citefmt{Kerscher \bgroup et al.\egroup
  }{1997b}]{kerscher:abell}
Kerscher, M., Schmalzing, J., Retzlaff, J., Borgani, S., Buchert, T.,
  Gottl{\"o}ber, S., M{\"u}ller, V., Plionis, M., \& Wagner, H.,
  {\mnras} \textbf{284}
  (1997), 73--84.

\bibitem[\protect\citefmt{Kogut \bgroup et al.\egroup }{1995}]{kogut:gaussian}
Kogut, A., Banday, A.~J., Bennett, C.~L., Hinshhaw, G., Lubin, P.~M., \& Smoot,
  G.~F., {\apjl} \textbf{439} (1995),
  L29--L32.

\bibitem[\protect\citefmt{Kogut \bgroup et al.\egroup }{1996}]{kogut:tests}
Kogut, A., Banday, A.~J., Bennett, C.~L., G{\'o}rski, K.~M., Hinshaw, G.,
  Smoot, G.~F., \& Wright, E.~L., {\apjl}
  \textbf{464} (1996), L29--L34.

\bibitem[\protect\citefmt{Luo \& Schramm}{1993}]{luo:kurtosis}
Luo, X. \& Schramm, D.~N., {\apj} \textbf{408}
  (1993), 33--42.

\bibitem[\protect\citefmt{Mecke \bgroup et al.\egroup }{1994}]{mecke:robust}
Mecke, K.~R., Buchert, T., \& Wagner, H., {\aanda} \textbf{288} (1994), 697--704.

\bibitem[\protect\citefmt{Minkowski}{1903}]{minkowski:volumen}
Minkowski, H., Mathematische Annalen \textbf{57} (1903), 447--495.

\bibitem[\protect\citefmt{Penzias \& Wilson}{1965}]{penzias:measurement}
Penzias, A.~A. \& Wilson, R.~W., {\apj}
  \textbf{142} (1965), 419--421.

\bibitem[\protect\citefmt{Pompilio \bgroup et al.\egroup
  }{1995}]{pompilio:multifractal}
Pompilio, M.~P., Bouchet, F.~R., Murante, G., \& Provenzale, A.,
  {\apj} \textbf{449} (1995), 1--8.

\bibitem[\protect\citefmt{Santal{\'o}}{1976}]{santalo:integralgeometry}
Santal{\'o}, L.~A., \emph{Integral geometry and geometric probability},
  Addison--Wesley, Reading, MA, 1976.

\bibitem[\protect\citefmt{Schmalzing \& Buchert}{1997}]{schmalzing:beyond}
Schmalzing, J. \& Buchert, T., {\apjl}
  \textbf{482} (1997), L1--L4.

\bibitem[\protect\citefmt{Schneider}{1978}]{schneider:curvature}
Schneider, R., Ann.\ Math.\ pura appl. \textbf{116} (1978), 101--134.

\bibitem[\protect\citefmt{Schneider}{1993}]{schneider:brunn}
Schneider, R., \emph{Convex bodies: the {B}runn--{M}inkowski theory}, Cambridge
  University Press, Cambridge, 1993.

\bibitem[\protect\citefmt{Smoot \bgroup et al.\egroup
  }{1994}]{smoot:topologyfirst}
Smoot, G.~F., Tenorio, L., J., B.~A., Kogut, A., Wright, E.~L., Hinshaw, G., \&
  Bennett, C.~L., {\apj} \textbf{437} (1994),
  1--11.

\bibitem[\protect\citefmt{Stebbins}{1988}]{stebbins:cosmic}
Stebbins, A., {\apj} \textbf{327} (1988),
  584--614.

\bibitem[\protect\citefmt{ter Haar~Romeny \bgroup et al.\egroup
  }{1991}]{koenderink:scalespace}
ter Haar~Romeny, B.~M., Florack, L.~M.~J., Koenderink, J.~J., \& Viergever,
  M.~A., in: \emph{Lecture Notes in Computer Science}, Vol. 511, Springer
  Verlag, Berlin, 1991, pp.~239--255.

\bibitem[\protect\citefmt{Tomita}{1986}]{tomita:curvature}
Tomita, H., Progr.\ Theor.\ Phys. \textbf{76} (1986), 952--955.

\bibitem[\protect\citefmt{Tomita}{1990}]{tomita:statistics}
Tomita, H., in: \emph{Formation, dynamics and statistics of patterns}
  (Kawasaki, K., Suzuki, M., \& Onuki, A., eds.), Vol.~1, World Scientific,
  1990, pp.~113--157.

\bibitem[\protect\citefmt{Torres \bgroup et al.\egroup }{1995}]{torres:genus}
Torres, S., Cay{\'o}n, L., {Mart{\'\i}nez--Gonz{\'a}lez}, E., \& Sanz, J.~L.,
  {\mnras} \textbf{274}
  (1995), 853--857.

\bibitem[\protect\citefmt{Torres}{1994}]{torres:topological}
Torres, S., {\apjl} \textbf{423} (1994),
  L9--L12.

\end{thebibliography}
\end{document}